\documentclass[useAMS,usenatbib]{mn2e}
\input psfig.sty

\def\eg{e.g.}
\def\LCDM{$\Lambda$CDM }

\def\c200{$c_{200}$}
\def\spin{$\lambda$}
\def\Section{\S}

\title[Concentrations of LSB Galaxy Haloes]
      {How Concentrated Are The Haloes Of Low Surface Brightness Galaxies In The Cold Dark Matter Model?}
      \author[Jeremy~Bailin, Chris~Power, Brad~K.~Gibson \& Matthias~Steinmetz]
	     {Jeremy~Bailin$^1$\thanks{Email: jbailin@astro.swin.edu.au},
	       Chris~Power$^1$, Brad~K.~Gibson$^1$ \& 
               Matthias~Steinmetz$^2$\\
	       $^1$ Centre for Astrophysics and Supercomputing,
               Swinburne University of Technology,
               PO Box 218, Hawthorn, 3122, Victoria, Australia\\
	       $^2$ Astrophysikalisches Institut Potsdam, An der Sternwarte 16,
	       D-14482 Potsdam, Germany
	     }

\begin{document}
	     
\date{submitted to MNRAS}
	     
\pagerange{\pageref{firstpage}--\pageref{lastpage}}\pubyear{2005}

\maketitle	     

\label{firstpage}
	     
\begin{abstract}
We investigate whether a correlation exists between the 
concentration (\c200) and spin (\spin) parameters of dark matter 
haloes forming in \LCDM\ N-body simulations. In
particular, we focus on haloes with virial masses in the range
$10^{11} \leq M_{\rm vir}/h^{-1} {\rm M_{\odot}} \leq 2\times 10^{12}$,
characteristic of the hosts of Low Surface Brightness (LSB) and
High Surface Brightness (HSB) galaxies. Our analysis demonstrates that 
\c200\ and \spin\ are anti-correlated. If we assume that
a galaxy disk forms in one of these haloes from baryons that
approximately conserved angular momentum during their dissipation into the
rotationally supported disk, then it is possible to estimate the disk's
central surface density. For an appropriate choice of mass-to-light
ratio, we can place constraints on the central surface brightness of
the disk and hence identify the analogues of the host haloes of LSB and
HSB galaxies. We find that our LSB galaxy analogues occupy haloes that have
lower concentrations than might be expected based on consideration of
the $M_{\rm vir}$--$c_{\rm vir}$ relation for the \LCDM cosmology. The 
distribution of concentrations peaks at \c200$\sim 6$, in good 
agreement with observational data, although there are important
differences between the shapes of the simulated and observationally inferred
distributions. This suggests that LSB galaxies inhabit a biased
subsample of the halo population, and that this bias may be an important
ingredient for resolving the current debate regarding the structure of
LSB galaxy dark matter haloes.
\end{abstract}


\setcounter{footnote}{1}

\section{Introduction}
\label{sec:introduction}

The Cold Dark Matter (CDM) model is widely accepted as the paradigm 
within which to understand the formation of structure in
the Universe. A defining characteristic of this model on the scale of galaxies
is the prediction that central densities of dark matter
haloes rise as $\rho \propto r^{\alpha}$, with $\alpha \approx -1$,
over the radii which govern the inner rotation curves of galaxies.
Over the last decade, numerous studies -- principally numerical -- have
sought to quantify and understand this divergent or \emph{cuspy} behaviour
\citep[\eg][]{nfw96,moore98,navarro04}. In particular, the continued assertion
that the rotation curves of dark matter dominated systems such as low surface 
brightness (LSB) galaxies require mass profiles with finite constant density 
cores or shallow ``cusps''
\citep[\eg][]{fp94,moore94,burkert95,deblok97,mcgaugh98,deblok03} has
prompted many detailed investigations into the consequences of these 
observations for the CDM model. 

This disparity between the predicted and inferred mass profiles appears to 
call into question the most basic assumption of the CDM model, namely the 
nature of the dark matter, and constitutes a critical challenge to the 
paradigm. Indeed, reconciling theory and observation has come to represent 
a central problem in contemporary cosmology and is arguably one of its most
contentious issues, as illustrated by recent analyses of the systematic 
effects that could affect high resolution rotation curve data of LSB galaxies 
and the fitting procedures used in mass modeling. 
\citet[][hereafter DB03]{deblok03} have argued
that the signature of CDM cusps should be clear in the data and so the 
model's predictions are inconsistent with observations, while 
\citet{hayashi04a,hayashi04b}, in some cases using overlapping data, find no 
compelling evidence for disagreement between cuspy CDM haloes and observed 
rotation curves. These studies focused on the shape of LSB galaxy
rotation curves, which rise approximately linearly with radius in the
manner one would expect if the mass density profile contained
only a shallow central cusp or a constant density core;
if the underlying profile is strongly cuspy, it is argued, the
rotation curve should rise more rapidly at small radii. However, the
rate at which the rotation velocity increases with radius is sensitive
to how centrally \emph{concentrated} the mass distribution within the
halo is -- the more centrally concentrated the halo, the more rapidly
its rotation curve will rise. Thus, a slowly rising rotation curve
could be interpreted as indicative of a halo that is weakly centrally 
concentrated.\\

The seminal work of \citet[][hereafter NFW]{nfw96} formalised this concept of
concentration. Based on a comprehensive suite of simulations, they 
established their \emph{universal} mass profile that could describe the 
structure of an average dark matter halo in dynamical equilibrium, independent
 of its mass or the choice of cosmological parameters and initial density 
perturbation spectrum. The circular velocity of a NFW halo can be expressed
as
\begin{equation}
\label{eq:nfw}
{\frac{V_c(x)}{V_{\rm vir}}={\sqrt{\frac{1}{x}\frac{\ln(1+c_{\rm vir}x)-c_{\rm vir}x/(1+c_{\rm vir}x)}{\ln(1+c_{\rm vir})-c_{\rm vir}/(1+c_{\rm vir})}}},}
\end{equation}
where $x=r/r_{\rm vir}$ is the radius $r$ normalised by the virial radius 
$r_{\rm vir}$\footnote{The virial radius defines the extent of a dark matter
halo and is such that it encloses an overdensity $\Delta_{\rm vir}$ times
the critical density of the Universe at a given epoch; for the $\Lambda$CDM
model, $\Delta_{\rm vir} \simeq 100$.}, $c_{\rm vir}=r_{\rm vir}/r_s$ is the 
\emph{concentration} parameter where $r_s$ is the scale radius, and 
$V_{\rm vir}={\sqrt{GM_{\rm vir}/r_{\rm vir}}}$ is the virial velocity of 
the halo. From equation~\ref{eq:nfw}, we observe that the circular velocity 
rises as $r^{1/2}$ at small radii, and the larger $c_{\rm vir}$, the greater
$V_c(x)/V_{\rm vir}$. 

NFW found that, on average, concentration decreased with increasing virial 
mass, and they understood that this reflected the mean density of the 
Universe at the time of the halo's formation; more massive systems collapse 
at later times when the mean density of the Universe is lower. \citet{nfw97} 
and subsequent studies by \citet[][hereafter B01]{bullock01} and 
\citet[][hereafter ENS]{eke01} formulated prescriptions that allow the 
concentration of a given halo of a given mass in a given cosmology to be 
predicted. Within the B01 or ENS formulations, we expect a typical galaxy 
mass halo ($2 \times 10^{12} M_{\odot}$) forming in the \LCDM cosmology
to have a concentration parameter $c_{\rm vir} \sim 10$.

If LSB galaxies occupy dark matter haloes of the kind that form in the \LCDM
model, then we might expect typical concentrations of $c_{\rm vir} \sim 10$.
Both \citet{mcgaugh03} and DB03 have considered the distribution
of concentrations obtained from fits to their LSB galaxy rotation curve data,
and as illustrated in Figure 3 of \citet{mcgaugh03} and Figure 1 of DB03, the 
distribution shows two well defined peaks, one at $c_{\rm vir} \sim 5$, the 
other at $c_{\rm vir}=0$.  The
preference for the data to favour such a low value of concentration appears
troubling for the CDM model.
The mass-concentration relation, predicts that $c_{\rm vir} \sim 5$
is typical of low mass clusters with $M_{\rm vir} \sim 10^{14} M_{\odot}$
rather than low mass galaxies, and the $1-\sigma$
scatter in concentration at a given
mass is expected to be $50\%$ at most.

Implicit in this comparison is the untested assumption that the haloes
of LSB galaxies are typical for their mass.
LSB galaxies are considered ideal for probing the structure of dark
matter haloes because the baryons contribute minimally to the rotation
curve, greatly facilitating the process of decomposing the rotation
curve into baryonic and halo contributions. However, if the haloes
of LSB galaxies are biased with respect to the general halo population,
then this is a biased comparison.
Indeed, it has been noted observationally
that detailed mass models of LSB galaxies reveal
dark matter haloes that are less dense and more extended than those of high 
surface brightness (HSB) counterparts of similar masses 
\citep[see the review of][]{bothun97}, suggesting lower concentrations than
might be expected. Furthermore, there are theoretical grounds to expect
that LSB galaxies inhabit a special class of haloes.
The low surface brightness of LSB galaxies is
a consequence of their low central density. If we make the reasonable 
assumption that specific angular momentum of the baryons is approximately 
conserved during their dissipation into a rotationally supported disk, then
the scale length of the resulting disk is related to the angular momentum of the
dark matter halo, and we anticipate that LSB galaxies will occupy dark matter 
haloes with larger spin parameters (a measure of the degree to which the halo
is supported by angular momentum; see \Section~\ref{sec:results}) than their
HSB counterparts. These considerations suggest that LSB galaxies occupy dark
matter haloes with low concentrations and high spin parameters, echoing the
conclusions of previous studies \citep[\eg][]{dalcanton97,jiminez98,mo98}.

In this paper, we investigate whether the correlation between concentration and
spin, as envisaged by \eg~\citet{dalcanton97},~\citet{jiminez98}, and 
\citet{mo98}, is sufficient to explain the distribution of
concentrations observed by \citet{mcgaugh03} and DB03 for their sample
of LSB galaxies. For this purpose, we have analysed 
a statistical sample of haloes, with masses in the range $10^{11} \leq 
M_{vir}/h^{-1}\, M_{\odot} \leq 2 \times 10^{12}$, forming in a high 
resolution cosmological \LCDM\ N-body simulation, and observe a 
trend for a halo's spin parameter to increase with decreasing concentration. 
Using the analytic prescription presented in \citet{mo98}, we use the halo 
spins to estimate disk scale lengths and surface densities for our 
hypothetical sample of galactic disks, and hence construct the distribution of
concentrations we might expect for the host haloes of LSB galaxies. Although
there are important differences between the distribution we recover and those
presented in \citet{mcgaugh03} and DB03, we
observe that the peak at $c_{\rm vir} \sim 5$ is reproduced in our 
distribution. We present the results of our analysis in the next section 
(\Section~\ref{sec:results}) and discuss their significance in
\Section~\ref{sec:discussion}.

\section{Results} 
\label{sec:results}

Our results are based on analysis of the $z$=0 output of a
$512^3$ particle cosmological N-body simulation of a 
periodic volume, $50~h^{-1}\, \rm Mpc$ on a side, that was run assuming
the $\Lambda$CDM cosmology ($\Omega_0=0.3$, $\Omega_{\Lambda}=0.7$, 
$\Omega_b=0.045$, $h=0.7$, $\sigma_8=0.9$) using the 
GADGET2 code (Springel et al., in preparation). A (comoving)
gravitational softening length of $5 h^{-1} \rm kpc$ (approximately
$1/20$th the mean interparticle separation) was adopted, and the chosen
parameters imply a particle mass of $7.75 \times 10^7 h^{-1}\,\rm
M_{\odot}$. The mass resolution of the simulation is sufficient to
resolve a $2\times 10^{12} h^{-1}\,\rm M_{\odot}$ halo with $\sim
26,000$ particles. Haloes are identified using a
\emph{friends-of-friends} \citep[FOF; \eg][]{davis85} algorithm with a linking
length of $b=0.2$ times the mean interparticle separation. We note that this
simulation formed the basis of recent studies by \citet{bs04a,bs04b}.\\

In order to facilitate comparison to observations,
we define the concentration with respect to
the radius that encloses a spherical overdensity of $200$ times the 
critical density (as in the original NFW papers), rather than the
cosmology dependent $\Delta_{\rm} \simeq 100$. This has the effect of
lowering the measured concentration of the halo by $\sim 30\%$
(approximately independent of concentration) relative to that predicted
by either the B01 or ENS prescriptions, reflecting the smaller measured
virial radius relative to a fixed scale radius. Concentrations are 
determined by identifying best fitting parameters ($r_{200}, c_{200}$) for
the enclosed mass at (0.06,0.12,0.25,0.4,0.6,1.0) $r_{\rm vir}$; we find
that this method is efficient and it produces accurate and robust
estimates of $c_{200}$.

The standard definition of the spin parameter,
\begin{equation}
\label{eq:lambda}
{\lambda = \frac{J |E|^{1/2}}{GM^{5/2}},}
\end{equation}
where $M,$ $J$ and $E$ are the mass, total angular momentum and energy of
the halo respectively, depends on binding energy, which is both a slow 
operation ($\sim N^2$) and ill defined in the context of a cosmological
simulation. Instead, we have chosen to compute the \citet{bullock01b} measure,
\begin{equation}
\label{eq:lambdaprime}
{\lambda'=\frac{J}{{\sqrt 2}MVR},}
\end{equation} 
where $V$ is the circular velocity at radius $R$; this can be corrected
by a factor $\lambda = \lambda' f(c_{200})^{1/2}$,
\begin{equation}
\label{eq:fc}
{f(c_{200})=2/3+(c_{200}/21.5)^{0.7} }
\end{equation}
\citep{mo98,bullock01b}.\\

\begin{figure}
 \psfig{figure=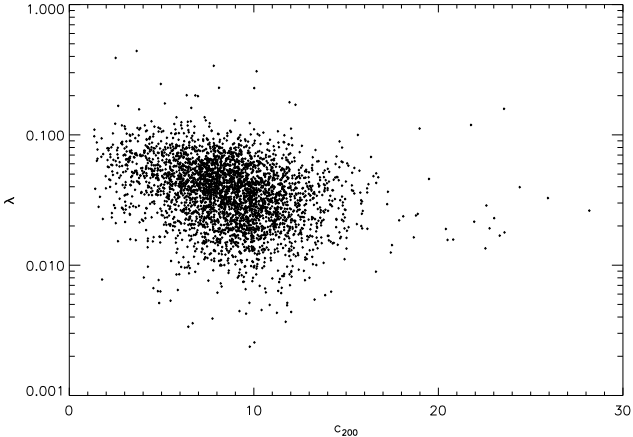,width=250pt}
  \caption{The correlation between spin parameter, $\lambda$, and 
concentration, $c_{200}$. We include all haloes with masses in the range
$10^{11} \leq M_{\rm vir}/h^{-1} {\rm M_{\odot}} \leq 2\times 10^{12}$. See 
text for further details.}
  \label{fig:lambdac}
\end{figure}

We begin by considering the distribution of halo concentrations $c_{200}$
relative to their spin parameters $\lambda$ for all FOF groups with masses 
between $10^{11}$ and $2\times 10^{12}\,h^{-1}\rm M_{\odot}$, as shown in
Figure~\ref{fig:lambdac}. Although there is large variation in
$\lambda$ at a given $c_{200}$, we note a discernible trend for $\lambda$ to 
increase with decreasing $c_{200}$; the median increases from
$\lambda_{\rm med} \simeq 0.03$ at $c_{200}=10$ to $\lambda_{\rm med}
\simeq 0.05$ at $c_{200}=5$. This is precisely the behaviour we would expect
the host haloes of LSB galaxies to display, in agreement with previous
work \citep[\eg][]{dalcanton97,jiminez98}.

Having established a correlation between $\lambda$ and
$c_{200}$, we wish to derive the quantity that characterises a LSB
galaxy, namely its central surface brightness $\mu_0$; our particular treatment
follows that of \citet{mo98}. If we assume that the angular momentum 
of the baryons is approximately conserved during their dissipation 
into a rotationally supported disk, we can estimate the angular
momentum of the disk $J_d$ and hence its extent for a given 
mass $M_d$. We assume that the disk's mass and angular momentum are 
fixed fractions of those of the host halo's mass and angular momentum, 
and choose $j_d=m_d=0.05$ such that $M_d=m_d\,M$ and $J_d=j_d\,J$. Accordingly,
the central surface density $\Sigma_0$ is given by
\begin{equation}
\label{eq:sigma0}
{\Sigma_0 = \frac{M_d}{2 \pi r_d^2},}
\end{equation}
where $r_d$ is the scale length of the disk,
\begin{equation}
\label{eq:rd}
r_d = \frac{r_{200}}{\sqrt{2}} \left(\frac{j_d}{m_d}\right) \lambda
  f(c_{200})^{-1/2} f_R(\lambda, c_{200}, m_d, j_d),
\end{equation}
where
\begin{equation}
\label{eq:fR}
f_R(\lambda, c_{200}, m_d, j_d) = 2 \left[ \int_0^\infty e^{-u}
  u^2 \frac{V_c(r_d u)}{V_{200}} \, \mathrm{d}u \right]^{-1},
\end{equation}
and we use the fitting formula of \citet{mo98} (see their equation (32)).
The central surface brightness $\mu_0$ can be estimated by assuming a
particular choice of local mass-to-light $(M/L)_{\rm tot}$ ratio, which
is a combination of the gas $(M/L)_{\rm HI}$ and star $(M/L)_{\ast}$
ratios. \citet{deblok96} indicates a local $(M/L)_{\rm HI} \sim 1$, and de
Blok (2004, private communication) suggests $(M/L)_{\ast} \sim 1$; in
what follows, we adopt $(M/L)_{\rm tot}=2$.

By definition, a LSB galaxy is one whose central surface brightness 
satisfies $\mu_0 \geq 23$ mag arcsec$^{-2}$; typical B-band values 
lie in the range $23 \leq \mu_0 \leq 24$ mag arcsec$^{-2}$ for LSB
galaxies, compared with $21 \leq \mu_0 \leq 22$ mag arcsec$^{-2}$ 
for HSB galaxies. Converting to central surface densities, we select 
LSB and HSB galaxies by applying cuts of $1.5 \leq \log \Sigma_0 \leq 1.9$
and $2.3 \leq \log \Sigma_0 \leq 2.7$ respectively, where
$\Sigma_0$ is expressed in units of $M_{\odot}~\mathrm{pc^{-2}}$.\\

\begin{figure}
  \psfig{figure=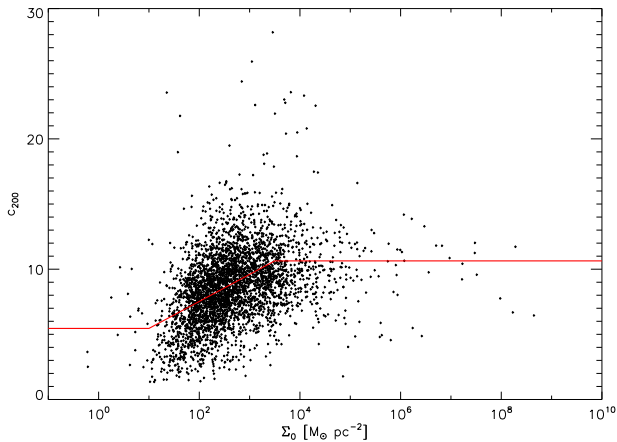,width=250pt}
  \caption{Correlation between central surface density, $\Sigma_0$, and
concentration, $c_{200}$. The solid line represents the
variation of the mean value of $c_{200}$ with $\Sigma_0$. As in 
Figure~\ref{fig:lambdac}, we consider only haloes with masses 
$10^{11} \leq M_{\rm vir}/h^{-1} {\rm M_{\odot}} \leq 2\times 10^{12}$.
See text for further details.}
  \label{fig:csig0}
\end{figure}

In Figure~\ref{fig:csig0}, we demonstrate how central surface density 
$\Sigma_0$ correlates with concentration $c_{200}$, where $\Sigma_0$ is
evaluated using equations~(\ref{eq:sigma0}) and (\ref{eq:rd}). As expected,
there is a clear trend for haloes with lower concentrations to host
disks that have lower central surface densities; the median increases from 
$\Sigma_{0,\rm med} \simeq 10^2~{\rm M}_{\odot}~{\rm pc}^{-2}$ at
$c_{200}=5$ to $\Sigma_{0,\rm med} \simeq 10^3~{\rm M}_{\odot}~{\rm
pc}^{-2}$ at $c_{200}=10$.
We quantify this trend by computing
the mean value of $c_{200}$ in bins of width 0.5 in $\log \Sigma_0$ and
performing a linear regression over the range $1 \leq \log \Sigma_0
\leq 3.5$ to obtain the relation,
\begin{equation}
\label{eq:csigma0}
{c_{200}=3.4+2.1 \log \Sigma_0,}
\end{equation}
shown by the solid curve in Figure~\ref{fig:csig0}; outside of this
range, $c_{200} \sim 5$ for $\log \Sigma_0 \leq 1$ and $c_{200} \sim
10$ for $\log \Sigma_0 \geq 3.5$.
This correlation is due in equal parts to the direct dependence of
$r_d$ on $c_{200}$ and on the indirect dependence via the spin
parameter \spin.
For a fixed $\Sigma_0$, we find the
distribution of $c_{200}$ has a width of $\sim 2.5-3$, essentially
independent of the precise value of $\Sigma_0$.\\

\begin{figure}
  \psfig{figure=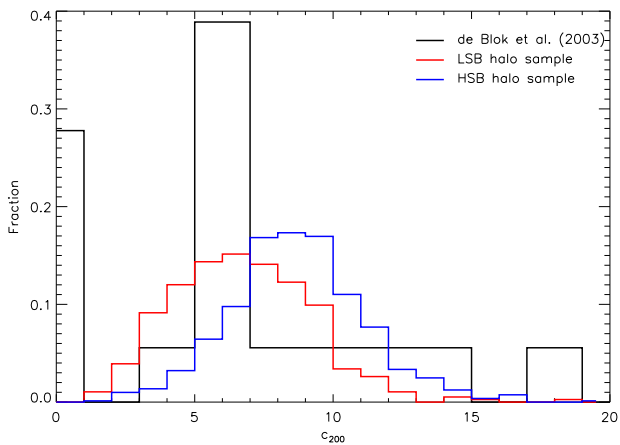,width=250pt}
  \caption{Distribution of concentrations. See text for further details.}
  \label{fig:chist}
\end{figure}

Finally, in Figure~\ref{fig:chist}, we show distributions of concentrations
corresponding to the host haloes of LSB (red histogram; hereafter the
LSB distribution) and HSB (blue histogram; hereafter the HSB
distribution) galaxies by applying the observationally motivated
cuts in central surface density ($1.5 \leq \log \Sigma_0 \leq 1.9$, 
$2.3 \leq \log \Sigma_0 \leq 2.7$ for LSB and HSB disks respectively) 
discussed above. For comparison, we show also the distribution derived
from fits to high resolution LSB galaxy rotation curves presented in
Figure 1 of DB03 (hereafter the DB03 distribution). 

There are two main points worthy of note in this figure. Firstly, we
find that the LSB and HSB distributions peak at different values of 
$c_{200}$; the peak value of the HSB distribution occurs at $c_{200} \sim 8.5$,
compared with $c_{200} \sim 6$ for the LSB distribution. Secondly, the 
values of $c_{200}$ at which the LSB and DB03 distributions peak
are in remarkably good agreement -- $c_{200} \sim 6$ -- although we
note that the overall shapes of the respective distributions differ;
the DB03 distribution has an additional well defined peak at
$c_{200}=0$ and a tail extending to $c_{200}=20$. 
However, the peak at $c_{200}=0$ in the DB03 distribution
is an artifact of the fitting procedure;
when the NFW fits were clearly unphysical, such as when they
preferred negative values of $c_{200}$, the value of $c_{200}$
was arbitrarily set to 0.1 \citep{dbb02}.
Therefore, these haloes should be characterized as haloes which cannot be
fit by the NFW form, rather than haloes with a well-defined concentration
of 0.1.
We also note that the $M_{\rm vir}$--$c_{200}$ relation implies that
very high concentration haloes are exclusively very low mass systems.
In particular, the observed halo with $c_{200} \sim 19$ is likely a dwarf
galaxy whose mass falls well below our low mass cutoff of
$10^{11}~h^{-1}~M_{\odot}$ \citep{thesis}.

We can estimate the likelihood that the DB03, LSB and HSB samples are
drawn from the same distribution by applying the K-S test. If we
include the entire DB03 sample, we find that it is inconsistent with
both the LSB ($P_{\rm KS}=0.04$) and HSB ($2\times 10^{-6}$) samples; 
ignoring the peak at $c_{200}=0$, we find reasonable agreement with the
LSB sample ($0.11$) but not the HSB sample ($0.002$); and 
finally, ignoring both the peak at $c_{200}=0$ as well as the 
high-$c_{200}$ tail, we find improved agreement with the LSB sample
($0.15$) but similarly poor agreement with the HSB sample ($0.001$).

\section{Discussion}
\label{sec:discussion}

The status of LSB galaxies as dark matter dominated systems is a well 
established observational result \citep[\eg][]{bothun97} and their 
rotation curves can provide, in principle, clean probes of the
distribution of dark matter within their host haloes. However, as the 
divergent conclusions reached by the recent studies of DB03
and \citet{hayashi04a} demonstrate, the question of whether the shape
of LSB galaxy rotation curves can be used to place strong constraints
on the nature of the dark matter and hence the validity of the \LCDM
model on small scales continues to be a matter of heated debate. 
We have not attempted to address this particular issue in our study -- 
the mass resolution of our simulation is insufficient to reliably
recover halo density profiles on the scales that are of most interest for 
comparison with observations \citep{power03} -- but it is
unlikely that their structure will differ significantly from that
predicted by NFW and more recent studies \citep[\eg][]{navarro04}.
Instead we have focused
on the related topic of the degree to which haloes of the kind
predicted to host LSB galaxies in the \LCDM cosmology are centrally 
concentrated.

Our analysis indicates that a correlation exists between concentration
$c_{200}$ and spin $\lambda$ for haloes with masses in the range we
have studied, $10^{11} \leq M_{\rm vir}/h^{-1} {\rm
  M_{\odot}} \leq 2\times 10^{12}$, such that haloes with low values of
$c_{200}$ tend to have high values of $\lambda$, and vice versa. By
assuming that the baryons associated with these low-$c_{200}$,
high-$\lambda$ haloes approximately conserve angular momentum during
the formation of the galaxy disk, we recast this trend as one
between the central surface density of the disk, $\Sigma_0$, and the
concentration of its host halo, $c_{200}$, using the formalism of
\citet{mo98}. Adopting an observationally motivated local mass-to-light
ratio, $(M/L)_{\rm tot}$, we identified those haloes most likely to
contain LSB galaxy disks, thus allowing us to construct a distribution
of concentrations that we compared with the distribution of inferred
concentrations presented in DB03. Although there are important
differences between the DB03 and our hypothetical LSB distributions, 
namely the presence of a significant number of galaxies for which
$c_{200}=0$ and several high-$c_{200}$ systems, we recover a peak value
of $c_{200} \sim 6$ that is in good agreement with observations.
In addition, when we exclude the $c_{200}=0$ systems, which are not
true measurements of halo concentration but rather a failing of the
entire fitting formula, and the $c_{200} \sim 19$ system, which likely
has a much lower mass than our haloes, we find that the observed and
predicted distributions of $c_{200}$ are consistent at the 15\%\ level.
\\

This is a compelling result because it establishes the relation between
a halo's concentration and its spin, and confirms the expectation that
haloes with low $c_{200}$ and high $\lambda$ are the natural hosts of
LSB galaxies. This is in good agreement with the observational
evidence, which indicates that LSB galaxies occupy dark matter haloes
that are more diffuse and more extended than those of HSB galaxies of 
comparable mass \citep[\eg][]{bothun97}. As noted in 
\S~\ref{sec:introduction}, this is equivalent to stating that the host
haloes of LSB galaxies are less centrally concentrated than those of their
HSB counterparts. B01 observed and quantified a scatter in concentration
for fixed halo mass, and found a 1-$\sigma$ variation of 
$\Delta\log c_{vir}\sim0.14$ (although we note that this has been
revised to $\Delta\log c_{vir}\sim0.18$; Bullock, private
communication). Thus we expect a galaxy halo with a virial mass 
$M_{200} = 2 \times 10^{12} h^{-1} {\rm M_{\odot}}$ to have a 
concentration in the range $6 \leq c_{200} \leq 12.5$ (B01, ENS).
However, previous studies failed to identify any correlation between a
halo's concentration and its spin; NFW examined the distribution of
residuals in characteristic overdensity $\delta_c$ (a function of the
concentration) and found no correlation with spin. Similarly, 
\citet{bullock01b} found no evidence for a correlation between 
the spin parameter $\lambda$
and concentration, although they noted a weak trend between their 
modified spin $\lambda'$ (see equation~\ref{eq:lambdaprime}) and
concentration $c_{\rm vir}$, such that smaller $c_{\rm vir}$
corresponded to larger $\lambda'$ and vice versa. We suggest that the 
absence of a correlation in the \citet{bullock01b} data reflects the
relatively poor mass resolution of their simulation in comparison with
ours (a factor of $\sim 10$ in mass), while we suspect that the
selective nature\footnote{NFW studied a biased sample -- systems in 
dynamical equilibrium.} and poor number
statistics of the NFW study will have made the correlation difficult to
identify.

Although this result does not resolve the current debate surrounding
the mass profiles of LSB galaxy haloes and the implications for the CDM
model, it is promising that a set of reasonable assumptions and a simple
correlation between concentration and spin can so elegantly lead to two
halo populations that bear the characteristics of LSB and HSB
galaxies. The central density profiles of the haloes must still be reconciled
with the detailed shapes of LSB rotation curves. One possibility is
that an additional correlation exists, such that LSB galaxies occupy 
low-$c_{200}$, high-$\lambda$ and \emph{triaxial} haloes, which may have 
observational consequences for the kind of rotation curves that we
might observe; indeed, the bearing of triaxiality on measured
rotation curves has been investigated by \citet{hayashi04b}.
Intriguingly, a subpopulation of our LSB host haloes have particularly
triaxial figures. Overall, we consider this an encouraging result for 
advocates of the CDM model and one worthy of further study.

\section*{Acknowledgements} 

We warmly thank Erwin de Blok and Eric Hayashi for helpful
correspondence. The financial support of the Australian Research
Council is gratefully acknowledged.

\label{lastpage}
\end{document}